\documentclass[8pt]{article}  \usepackage{times}
\pdfoutput=1
\usepackage{graphicx}

\usepackage{color}

\begin{document}

\twocolumn[{\LARGE \textbf{The thermodynamics of lipid ion channel formation in the absence and presence of anesthetics. BLM experiments and simulations.\\*[0.0cm]}}

{\large Katarzyna Wodzinska$^{\dagger}$, Andreas Blicher$^{\dagger}$,  and Thomas Heimburg$^{\ast}$\\*[0.1cm]
{\small Niels Bohr Institute, University of Copenhagen, Blegdamsvej 17, 2100 Copenhagen \O, Denmark\\}

{\normalsize \textbf{ABSTRACT\hspace{0.5cm} It is known that lipid membranes become permeable in their melting regime. In microscopic conductance measurements on black lipid membranes one finds that conduction takes place via quantized events closely resembling those reported for protein ion channels. Here, we present data of ion currents through black lipid membranes in the presence and absence of the anesthetics octanol and ethanol, and compare them to a statistical thermodynamics model using parameters that are obtained from experimental calorimetric data. The conductance steps in pure lipid membrane suggest aqueous pores with the size of approximately one lipid cross-section.  We model the permeability by assuming empty sites of the size of one lipid. We find that pore formation in the melting transition regime is facilitated by the increase of the lateral compressibility that expresses itself in the area fluctuations.  Thus, pore formation is related to critical opalescence in two dimensions. Anesthetics alter the permeability by affecting the thermodynamic state of the membrane and by shifting the heat capacity profiles.}\\*[0.0cm] }}
]
\setlength{\parindent}{0cm}
\footnotesize {$^{\dagger}$K. Wodzinska and A.Blicher contributed equally to this work. \\
$^{\ast}$corresponding author, theimbu@nbi.dk}\\

\setlength{\parindent}{0cm}
\footnotesize{\textbf{Abbreviations:} DSC, differential scanning calorimetry; LUV, large unilamellar vesicle; }
\setlength{\parindent}{0cm}
\footnotesize{BLM, black lipid membrane;}
\footnotesize{DMPC, 1,2-dimyristoyl-sn-glycero-3-phosphocholine;}
\footnotesize{DOPC, 1,2-dioleoyl-sn-glycero-3-phosphocholine;}
\footnotesize{DPPC, 1,2-dipalmitoyl-sn-glycero-3-phosphocholine;}
\footnotesize{DSPC, 1,2-distearoyl-sn-\linebreak glycero-3-phosphocholine;}
\footnotesize{MC, Monte Carlo;}

\normalsize

\section{Introduction}\label{intro}
In biology, lipid membranes are typically considered as insulators, which is an important assumption for models related to nerve pulse propagation (e.g., the Hodgkin-Huxley model \cite{Hodgkin1952}) and the conduction of ions through proteins. It is known, however, that lipid membranes become permeable in the proximity of membrane melting transitions \cite{Papahadjopoulos1973, Nagle1978b, Corvera1992, Blicher2009}, which creates significant conceptual problems for such models because biological membranes are in fact often close to such transitions under physiological conditions \cite{Heimburg2005c}. The increased permeability of membranes in transitions has been attributed to the large fluctuations in membrane area generally related to transitions, and
to the appearance of domain boundaries \cite{Papahadjopoulos1973, Cruzeiro1988}. The area fluctuations are proportional to the lateral compressibility  \cite{Heimburg1998} and the compressibility is therefore high in the transition regime. Thus, the work necessary to create conducting membrane defects under these conditions is small \cite{Nagle1978b}. \\
Interestingly, it has been found in microscopic measurements using black lipid membranes or using patch-clamp that the conductance of ions through pure lipid membranes occurs in quantized steps, suggesting pores of well-defined size \cite{Yafuso1974, Antonov1980, Kaufmann1983a, Kaufmann1983b, Antonov1985, Antonov2005, Blicher2009}. These events display a similar conductance and typical opening and closing time scales as the conduction events reported for protein ion channels \cite{Hille1992}. In a recent paper we found conductance steps suggesting aqueous pores of 0.7 nm, very similar to the size proposed for protein ion channels \cite{Blicher2009}. Pores of this size have also been proposed on the basis of molecular dynamics simulations \cite{Boeckmann2008}. The obvious question arises how the lipid conduction events can be distinguished from those of proteins. In particular, while lipid membranes can be investigated in the absence of proteins, channel proteins cannot be investigated in the absence of lipid membranes into which they are embedded. A patch pipette has a diameter of about 1 $\mu$m while the diameter of a typical membrane protein is of the order of 5 nm. The patch tip area is therefore about 40000 times larger than the protein, which renders it technically impossible to measure currents through proteins without simultaneously measuring the events in the lipid membrane, too. \\
Recently, using fluorescence correlation spectroscopy we \linebreak showed empirically that the permeability of membranes is with\-in experimentally accuracy proportional to the heat capacity in the transition range \cite{Blicher2009}. Thus, everything that shifts the heat capacity profile of membrane melting will also alter the membrane permeability. This includes changes in temperature, pressure and pH, but also the presence of anesthetics. The latter molecules have been shown to lower membrane transitions according a freezing point depression law, completely   independent of the nature of the anesthetics drug \cite{Heimburg2007c}. When measuring in the transition regime, anesthetics therefore lower the permeability, and quantized current events are blocked \cite{Blicher2009}

In this paper we show experimental data on quantized currents through black lipid membranes (BLMs) and the effect of anesthetics on the conduction. We rationalize our findings with a statistical thermodynamics model that takes the area fluctuations of the membranes into account. Statistical thermodynamics models (e.g., 2 state Ising models or 10 state Pink models) in the past have been successfully used to describe cooperative melting events in membranes \cite{Doniach1978, Mouritsen1983, Sugar1994, Heimburg1996a, Sugar1999, Ivanova2001, Ivanova2003}. In transitions, the cooperative fluctuations lead to domain formation on scales much larger than the individual lipid. On these scales, the lack of molecular detail is not important. For this reason,. Here, we allow for the formation of pores facilitated by the fluctuations in area. We demonstrate that this models results in channel statistics similar to the BLM experiment and to explain the effect of anesthetics.

\section{Methods}\label{Methods}

\subsection{Materials}\label{Materials}
\noindent\textbf{Chemicals.}
Decane and 1-octanol were purchased from Fluka Chemie AG (Deisenhofen, Germany), N-hexadecane, chloroform and ethanol were obtained from Merck (Hohenbrunn, Germany), n-pentane was provided by BDH (Poole, UK) and potassium chloride from J.T. Baker Analyzed (Deventer, Holland). 1,2-dipalmitoyl-\-sn-\-glycero-3- phosphocholine (DPPC) and 1,2-\-dioleoyl-\-sn-\-glycero-3-\-phosphocholine (DOPC) were purchased from Avanti Polar Lipids (Birmingham, AL) and used without further purification. For all experiments, MilliQ water (18.1 M$\Omega$) was used.\\
\noindent\textbf{Calorimetry.}
DSC experiments were performed on a VP-DSC Calorimeter (MicroCal, Nort{-}hampton, MA, USA) with a scan rate of $5^{\circ}$C/h. The lipid samples were prepared by pre-dissol\-ving in chloroform, which was first dried under a nitrogen \linebreak stream and the kept under high vacuum overnight. The dried lipid mixtures were dispersed in MilliQ water to a final concentration of 20mM.  The buffer used was the same as for the BLM experiments. Before filling the calorimeter, the solutions were degassed for 10 minutes.\\
\noindent\textbf{Black lipid membranes.}
Planar bilayers were formed over a round aperture of $\approx$80 $\mu$m radius in a Teflon film of 25$\mu$m thickness dividing two compartments of a teflon chamber embedded in a brass block that could be heated by a circulating water bath. The aperture in the Teflon film was prepainted with 5\% hexadecane in pentane. The BLMs were painted with DOPC:DPPC 2:1 lipid solutions in decane/chloroform/metha\-nol 7:2:1 and formed following the method described by Montal and M\"uller \cite{Montal1972}. The two compartments of the Teflon block were filled with unbuffered 150mM KCl (pH~6.5). Lipid solution (25mg/ml) was spread on the buffer surface in each compartment (approx.\ 3 $\mu$l on each side). Ag/AgCl electrodes were placed into both compartments of the chamber. After 15-30 minutes to allow for the evaporation of the solvent, the water level of the compartments was lowered and raised several times until a bilayer was formed over the hole. The formation of a BLMs was controlled visually and by capacitance measurements (with triangular 100mV voltage input pulse). For the experiments with octanol, 15\% v/v octanol in methanol solution was prepared and added symmetrically on both sides of the Teflon chamber. To calculate the concentration of octanol in the membrane we assumed an octanol partition coefficient between lipid membrane and water of 200 \cite{Jain1978}. In the experiments with ethanol we used 60 $\mu$l and 120$\mu$l of 99\% ethanol for a buffer volume of the two compartments of 2ml. Assuming a ethanol partition coefficient of 0.48 between membranes and water  \cite{Firestone1986} this corresponds to 9.5 mol\% and 19 mol \% ethanol in the membrane.\\
Conductance measurements were performed on an Axopatch 200B amplifier in voltage clamp mode connected to a DigiData 1200 digitizer (both Molecular Devices, Sunnyvale, CA, USA). Current traces were filtered with 1kHz low-pass Bessel filter and recorded with Clampex 9.2 software (Axon Instruments) on the hard drive of the computer using an AD converter with a time resolution of 0.1ms. The data was further analyzed with Clampfit 9.2 and low-pass filtered with Bessel (8-pole) filter at a cut-off frequency of 300Hz. Temperature was controlled by a HAAKE DC30 K20 (Waltham, USA) waterbath and a thermocouple (WSE, Thermocoax).\\


\subsection{Simulating pore formation}\label{MC} 

Here we introduce a statistical mechanical model for the description of pore formation close to lipid melting transitions in a system with one species of lipids and a second species of molecules (anesthetics), based on previous work by \cite{Ivanova2003}. In this work each lipid could be in either gel or fluid state. This model formally corresponds to an Ising model in a field. In the present work we additionally allowed for the formation of pores in the membrane by moving lipids from randomly chosen lattice sites to the edge of the computer matrix. For a given temperature we evaluate the mean number of gel and fluid lipids and the fluctuations in their number, and we determine the number of spontaneously formed pores. 
\begin{figure*}[ht!]
    \begin{center}
	\includegraphics[width=16.5cm]{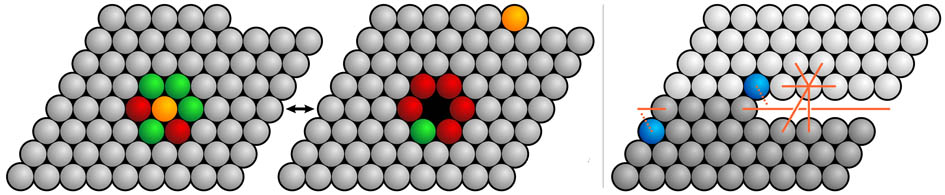}
	\parbox[c]{16cm}{\caption{\small\textit{Left and center: The pore formation step. Green represents fluid state lipids, while the gels are red. The yellow particle can be either a lipid or an anesthetic molecule. In this Monte Carlo step three lipids simultaneously change state while the yellow particle is moved to or from  the edge of the lattice. This means that the pore formation step increases the lattice size by one. During this step the overall area of the matrix stays constant (see text). Right: The dark and light grey regions (bottom and top half) indicate the boundaries of repeating lattices. The definition of the nearest neighbors of the particle moved from the site of the pore to the edge. The red lines indicate the definition of nearest neighbors at the edge. Due to the odd shape of the lattice the blue lattice sites display a different definition of nearest neighborhood (indicated by the dotted red line). Still all lattice sites have 6 nearest neighbors. }
	\label{Figure1}}}
    \end{center}
\end{figure*}

A numerical evaluation by means of Monte Carlo simulations was used. To do this one first has to define the various Monte Carlo steps. Whether a given step is accepted or not, depends on difference in the free energy of the old and the new system configuration as given by eq. (1). At its most basic (i.e. a one-component system without anesthetics and no pore formation) only one type of step is needed, namely one that allows the lipids to change their state (a "melting step"). As we are using the two-state Ising model as a basis, each lipid is assumed to be in either an ordered (gel) or disordered (fluid) state \cite{Heimburg1996a, Ivanova2003}. For the more advanced models (multiple lipid species, anesthetics, pore formation, etc.) it is also necessary to include the possibility for any two particles to swap positions (i.e., a "diffusion step", \cite{Hac2005,  Seeger2005}), as well as to include a Monte Carlo step which allows for the creation/sealing of a pore. The basic idea behind the pore forming step is the observation that a lipid changes area by approximately 25\% when it goes from the fluid to the gel state. So if three neighboring fluid lipids simultaneously change states, one can create a pore which has an area that is equal to a gel state lipid. Thus this step both conserves the number of lipids and also the area as $A_{before}=3A_{fluid}$$=A_{after}=3A_{gel}+A_{pore}$. In this kind of system a pore has automatically the cross-section of one lipid, which is what is observed in experiments \cite{Blicher2009} and molecular dynamics simulations \cite{Boeckmann2008}.\\
First pore formation is described. This step can be applied if a matrix site (that is not already a pore) is randomly picked: first the surroundings (nearest neighbors) is  checked for lipids in the fluid state.  If there are three or more, a pore is temporarily created by  changing the three fluids into gels, and moving the  "central particle" to the "end" of the lattice,  increasing the number of lattice sites by one  (see Fig.~\ref{Figure1}, left and center). In order to obey detailed balance, there must also be a possibility to seal a pore. This can happen when a pore is picked,  and it has three or more neighbors which are gel state lipids.  If so, the pore is temporarily sealed by changing the three gels to fluids,  and moving the "end particle" into the pore, thus reducing the number of lattice sites by one (see Fig.~\ref{Figure1}). Finally, the proposed change is accepted or rejected using the Glauber algorithm \cite{Glauber1963}. We used periodic boundary conditions using a special definition of the nearest neighbor lipid environment on the edge (see Fig.~\ref{Figure1}, right).

\textbf{Model parameters} \label{determination_of_model_parameters}

There are three parameters which need to be determined if one wants to describe the melting transition of the most basic one-component two-state lipid system, namely the melting enthalpy, $\Delta H$, the melting entropy, $\Delta S$, and the cooperativity parameter, $\omega_{fg}$. The latter parameter describes the interaction between a gel and a fluid lipid and determines the half width of the melting transition. For the model additionally containing anesthetics and pores, another five interaction parameters are needed, namely $\omega_{af}$ (anesthetic/fluid), $\omega_{ag}$ (anesthetic/gel),  $\omega_{ap}$ (anesthetic/pore), $\omega_{pg}$ (pore/gel),  and $\omega_{pf}$ (pore/fluid) (\cite{Ivanova2001}).

With these parameter given the changes in free energy can be calculated as
\begin{eqnarray}
\Delta G =&& \Delta H - T \Delta S + \Delta N_{fg} \omega_{fg} \nonumber\\
  &+& \Delta N_{af} \omega_{af}  + \Delta N_{ag} \omega_{ag}  \\
  &+& \Delta N_{pf} \omega_{pf} + \Delta N_{pg} \omega_{pg} 
  + \Delta N_{ap} \omega_{ap}\nonumber
 \label{deltaG}
\end{eqnarray}

Fortunately, the first three parameters can be obtained from calorimetric  experiments (\cite{Ivanova2003}), while the remaining three will need to be guesstimated from permeation measurements.  Specifically, one can obtain the melting enthalpy and entropy of the lipids from the heat capacity profile of the one-component lipid system: the enthalpy change of the transition is obtained by integrating over the excess heat capacity in the relevant temperature interval. The entropy change can then easily be determined via $\Delta S = \Delta H / T_{m}$, as the transition midpoint, $T_{m}$, can be read off the heat capacity profile directly. The cooperativity parameter, $\omega_{fg}$, relates directly to the width of the transition peak, and  can consequently be determined by fitting the simulated heat capacity profiles to the experimentally measured one. 

The five remaining interaction parameters, $\omega_{ij}$, can be obtained indirectly from various other experiments. We have shown before that most anesthetics generate a freezing point depression that is consistent with ideal solubility in the fluid phase and no solubility in the gel phase \cite{Heimburg2007c}. Therefore we chosen $\omega_{af}=0$, and a high value for the anesthetic-gel interaction with $\omega_{ag}=\omega_{fg}$ (\cite{Ivanova2003}). Determination of the remaining three parameters ($\omega_{ap}$, $\omega_{pg}$  and $\omega_{pf}$) can, in principle, be done by comparing the simulated permeability (basically the average number of pores) to measurements. \\
Even before doing this, it is possible to predict a number of relations,  if we assume that hydrophobic matching is a major determinant of  the nearest neighbor interactions.  Firstly, if the added anaesthetic is strongly hydrophobic, it seems  reasonable to assume that $\omega_{ap} \gg \omega_{fg}$, as the  cooperativity parameter, $\omega_{fg}$, is largely determined by the hydrophobic mismatch between a  fluid and gel state lipid. Similarly, this should mean that $\omega_{pg} > \omega_{pf}$ for two reasons: 1) the lipid in the gel state would have a larger part of the hydrophobic chains exposed to the water, and 2) the fluid phase is more loosely packed, meaning that defects and spaces between the lipids (i.e. pores) would be more likely to appear than in the tightly packed gel phase. 

Based on the considerations above, the parameter values used in the simulations were as follows:
$\Delta H = 36400$ J/mol, 
$\Delta S = 115.87$ J/(mol$\cdot$K),
$\omega_{fg} =	1326$ J/mol,  
$\omega_{pf} =	 3700$ J/mol,  
$\omega_{pg} = 6000$ J/mol,  
$\omega_{pa} = 13260$ J/mol,   
$\omega_{af} =	 0.0$ J/mol, and
$\omega_{ag} = 1326$ J/mol.
The values for $\Delta H$, $\Delta S$ and $\omega_{fg}$ are identical to those used in \cite{Ivanova2003}. 
As seen in the results part, the average number of pores form in the BLM experiments is in the range of 1-2 in an area of 50$\mu$m$^2$ corresponding to 10$^8$ lipid molecules. In our simulations we used a matrix with 10000 lipids that is way smaller than the experimental membrane. Therefore, for the pore formation we opted for a set of pore-lipid interaction parameters which on average generates a similar number of pores as in the experiment requiring smaller pore-lipid interaction parameters as one would expect in a natural membrane. Additionally, it was necessary to disallow direct pore-pore interactions, so as to avoid run-away aggregation of pores and infinite growth of the simulation matrix. The effect on the results should be negligible, as typically the pore density in the simulation was so low as to make pore-pore interactions insignificant. In the experiment collision of pores is much less likely because of the much larger area, but may occur under stress when the membrane ruptures. The exact appearance of the pores vs. temperature profiles will, of course, depend on the choice of parameters. However, for a large range of interaction values, the permeability profile shows a strong peak at the phase transition temperature, and the position of this  peak was found to always follow the peak of the heat capacity profile. 
Typical system sizes were $10^4$ particles on a triangular lattice with periodic boundary conditions. The typical simulation was allowed to equilibrate for $2\cdot 10^4$ to $10^5$ Monte Carlo cycles (lattice sweeps) two times -- first without pores involved, and then again with pore formation allowed. The actual sampling was then done over another $10^5$ to $5\cdot 10^5$ cycles. Most of the calculated curves presented were in addition averaged over several independent runs.

%

\section{Results}\label{Results}
In this paper we investigate pore formation in lipid membranes both experimentally with black lipid membranes (BLM) using the Montal-M\"uller technique \cite{Montal1972}, and theoretically with Monte Carlo simulations. The goal is to demonstrate that pore formation is related to the heat capacity and the cooperative area fluctuations, respectively, and that anesthetics change the permeability in a coherent manner.
\begin{figure}[htb!]
    \begin{center}
	\includegraphics[width=8.5cm]{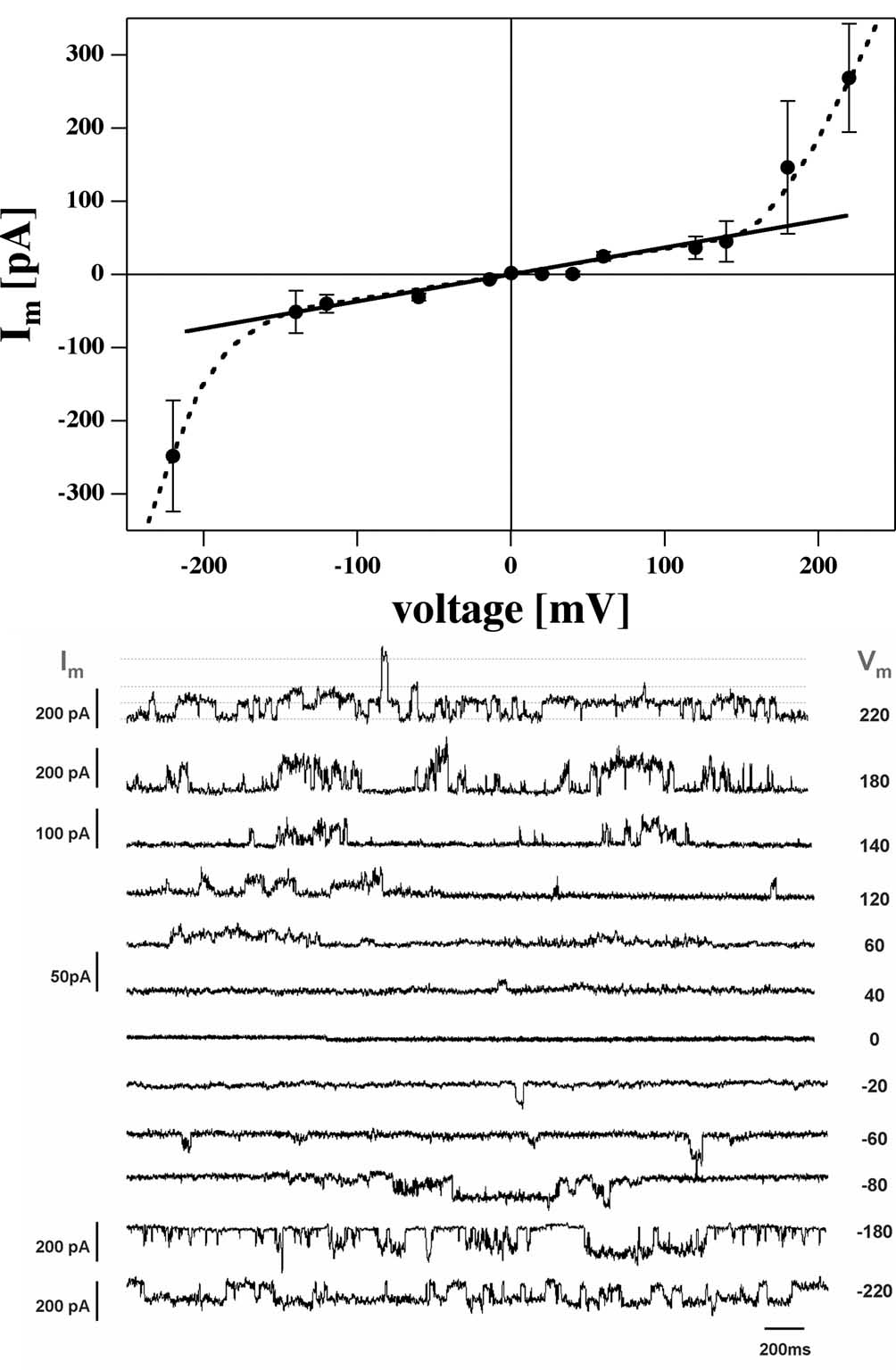}
	\parbox[c]{8cm}{ \caption{\small\textit{Bottom: Current traces through BLMs of a DOPC:DPPC=2:1 mixture  (150mM KCl, pH$\approx$6.5 at T=19$^{\circ}$C)  for different voltages. Top: Current-voltage relationship for the traces from the bottom panel. One finds a linear current-voltage relationship in the range from -150mV to +150mV. At higher voltages the traces become nonlinear. A total of 328 current traces with 30 seconds each where used for this graph.}
	\label{Figure2}}}
    \end{center}
\end{figure}

\subsection{Black lipid membranes}\label{Results_BLM}
\begin{figure*}[ht!]
    \begin{center}
	\includegraphics[width=16.5cm]{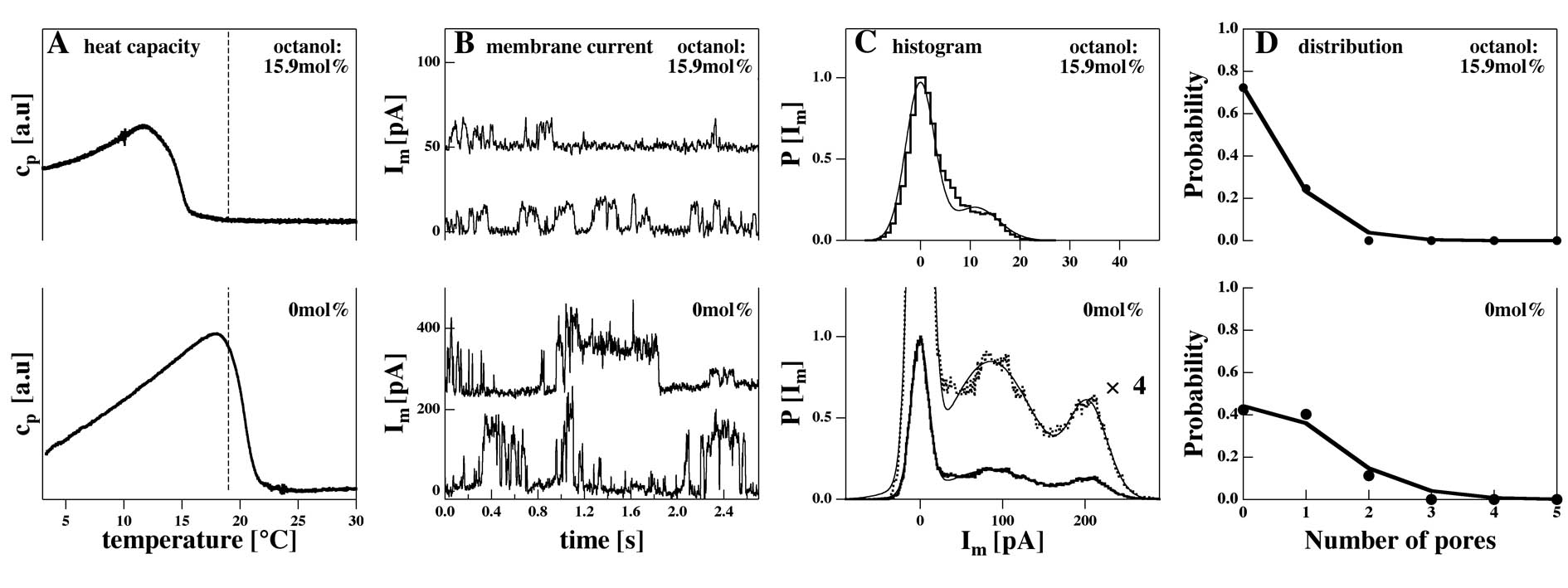}
	\parbox[c]{16cm}{\caption{\small\textit{Experiments on BLMs of a DOPC:DPPC=2:1 mixture  (150mM KCl, pH$\approx$6.5) in the presence (top) and the absence (bottom) of 15.9 mol\% octanol in the membrane. A: Heat capacity profiles. The dashed line indicates experimental temperature. Octanol shifts the calorimetric events towards lower temperatures. In the presence of octanol the experimental temperature above the melting events while it is at the $c_p$-maximum in the absence of octanol. B: Current traces obtained at T=19$^{\circ}$C and 210 mV. Two representative traces are given for the two experimental conditions, respectively. Note the different amplitudes of the currents and the altered frequencies of current events. C: Current histograms from the curves in panel B showing on or two current events. D: Analysis of the frequencies of the number of pores. Symbols represent the areas of the peaks in panel C. The solid line represents a best fit using a Poisson-distribution. }
	\label{Figure3a}}}
    \end{center}
\end{figure*}

In the following we show the conduction events through synthetic BLMs. We have chosen to work with lipid mixtures that have their transition events close to room temperature and that display broad heat capacity profiles. The latter makes it easier in experiments to adjust the temperature such that one is very close to the heat capacity maximum. \\
Fig. \ref{Figure2} shows the current traces for a DOPC: DPPC=2:1 mixture at 19$^{\circ}$C (150mM KCl, pH $\approx$ 6.5) for various transmembrane voltages between -220mV and +220mV. The heat capacity profile for this lipid mixture can be found in Fig. \ref{Figure3a}A (bottom). It displays a $c_p$-maximum close to 19$^{\circ}$C.  In agreement with earlier publications (e.g., \cite{Antonov1980, Blicher2009} we find quantized conduction events very similar to those of reported for protein ion channels (Fig. \ref{Figure2}, bottom) with voltage dependent amplitudes.  The amplitudes display a linear dependence of the transmembrane voltage, $V_m$, between about -150 to + 150 mV (Fig. \ref{Figure2}, top). At voltages above $|V_m|$= 150 mV, the conductance is found to be higher than expected from the linear relation.  The origin of this non-linear current-voltage relationship is not clear but may be related to the influence of voltage on the phase behavior of the lipid membranes. It has been shown by other authors that the position of the melting transition can be influenced by voltage \cite{Heckl1988, Antonov1990, Lee1995}. Thus, we suspect that the nonlinear behavior at higher voltages is a consequence of the influence of voltage on the phase transition.\\

\begin{figure*}[htb!]
    \begin{center}
	\includegraphics[width=16.5cm]{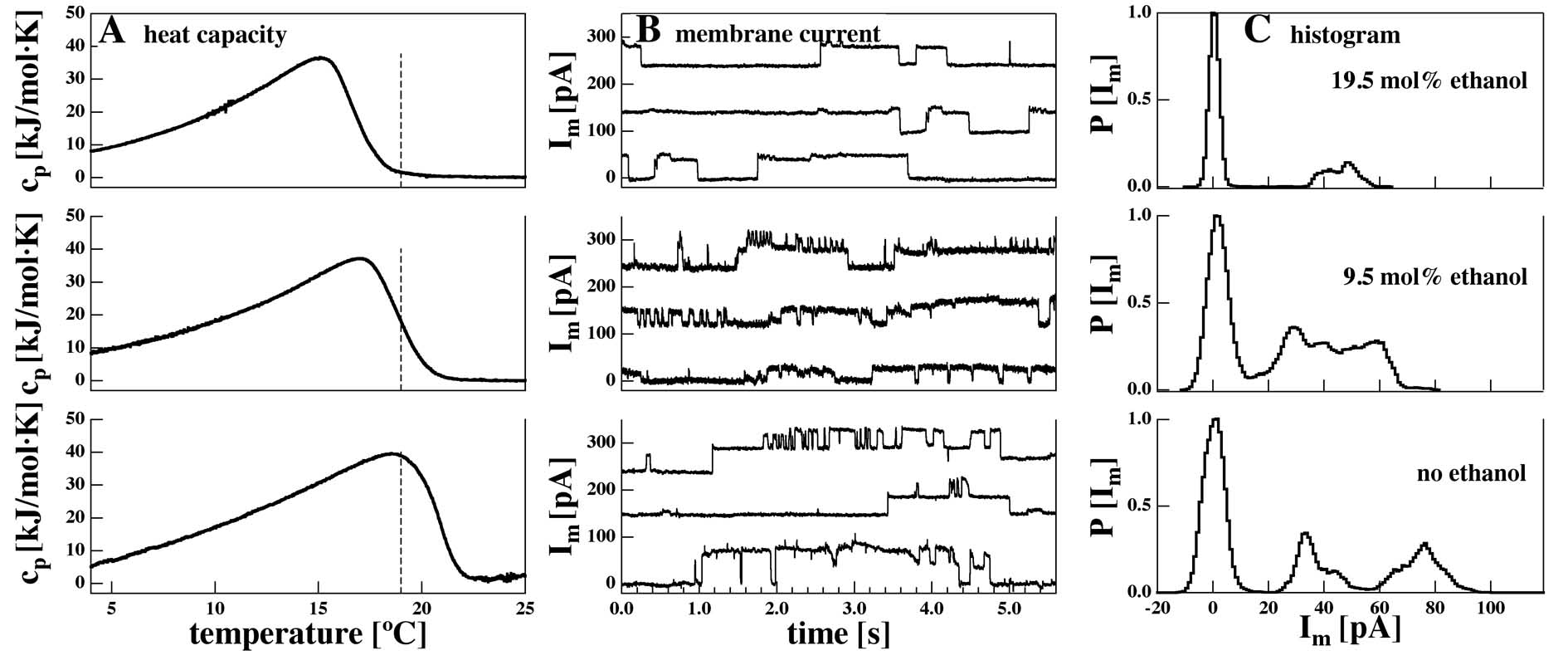}
	\parbox[c]{16cm}{\caption{\small\textit{Experiments on BLMs of a DOPC:DPPC=2:1 mixture  (150mM KCl, pH$\approx$6.5) in the presence of 19.5 mol\% ethanol (top), 9.5 mol\% (middle) and no ethanol in the fluid lipid membrane. A: Heat capacity traces.  B: Current traces measured at 120 mV. The presence of ethanol seems to generate longer opening times. Three representative traces are given for each experimental condition. C: Histograms of the current events. The behavior is different as in the presence of octanol (Fig. 3] ). In the absence of ethanal one find two clearly visible equally spaced current steps of about 39pA that merges into one peak in the presence of ethanol. }
	\label{Figure3b}}}
    \end{center}
\end{figure*}

\textbf{The effect of anesthetics}
Fig. \ref{Figure3a} shows the influence of octanol on the permeation of the above membranes measured at 19$^{\circ}$C and 210 mV. The top panels show heat capacity profiles (Fig. \ref{Figure3a}A), representative current traces (Fig. \ref{Figure3a}B), current histograms (Fig. \ref{Figure3a}C) and pore number probability distribution (Fig. \ref{Figure3a}D) in the presence of 15.9 mol\% octanol  in the membrane (calculated using an octanol partition coefficient in the membrane of 200, \cite{Jain1978}). The bottom panels show the respective data in the absence of octanol. One can see that the presence of octanol shifts the calorimetric profiles towards lower temperatures such that one is above the chain melting regime in the presence of octanol and in the transition regime in the absence of octanol. The current traces display smaller current amplitudes and a lower frequency of current events in the presence of octanol. Fig. \ref{Figure3a}C shows current amplitude histograms obtained from the data shown in Fig. \ref{Figure3a}B. In the absence of octanol one finds at least two current steps corresponding to the two maxima of the histogram that are different from zero. In the presence of 15.9 mol\% octanol one finds only one current step with a much lower amplitude. Fig. \ref{Figure3a}D shows the areas of the current peaks from Fig. \ref{Figure3a}C (obtained from fitting the peaks with Gaussian profiles). The solid lines are fits to a Poissonian distribution. Such distributions are expected for a low mean number of lipid ion channels. \\
Fig. \ref{Figure3b} shows the corresponding experiments in the presence of ethanol. Ethanol is a very weak anesthetic with a membrane-water partition coefficient of about 0.48 \cite{Firestone1986} (octanol has about 200). Three experiments recorded at $V_m$=120mV are shown, with 0 mol\%, 9.5 mol\% and 19.5 mol\% ethanol in the fluid lipid membrane. Increasing amounts of ethanol shift the melting profile towards lower temperatures. The corresponding current traces show at least two current steps at 39 and 78 pA in the absence of ethanol (note that the voltage is lower than in the octanol experiment in Fig. \ref{Figure3a}), while there seems to be only one step for 19.5 mol\% of  around 45pA. The current histograms suggests that the two currents steps seen in the absence of octanol merge into one in the presence of ethanol. This behavior is different from that of octanol where the overall conductance decreased upon the addition of anesthetics. We did not attempt to make a Poisson analysis of the histograms of the ethanol containing membranes since the histograms in the presence of ethanol do not contain evenly spaced conduction levels in the presence of ethanol.\\ 
In the discussion section we show that the response of the lipid ion channels to both octanol and ethanol is very similar to that reported for  protein ion channels.\\

\subsection{Monte Carlo simulations}\label{Results_MC}

In the following we attempt to rationalize the experimental findings obtained from anesthetics-containing membranes by a statistical thermodynamics model that is describe in detail in the Materials section. The main idea in this simulation is to postulate a pore formation process that conserves the overall area of the complete membranes. The conservation of area during the pore formation step implies that pores can only form when compressing the area of the lipids. Hence, the pore formation process is linked to the area fluctuations of the lipid matrix. The area fluctuations, however, are proportional to the lateral compressibility. Nagle and Scott \cite{Nagle1978b} argued that the likelihood of finding a pore is proportional to the compressibility. Now it is known that the excess heat capacity of lipid melting is proportional to the compressibility \cite{Heimburg1998}. We have therefore recently argued that the likelihood to find a pore must be intimately related to the heat capacity \cite{Blicher2009}. Our Monte Carlo model makes use of the considerations of Nagle and Scott \cite{Nagle1978b}.\\
\begin{figure}[hb!]
    \begin{center}
	\includegraphics[width=8.5cm]{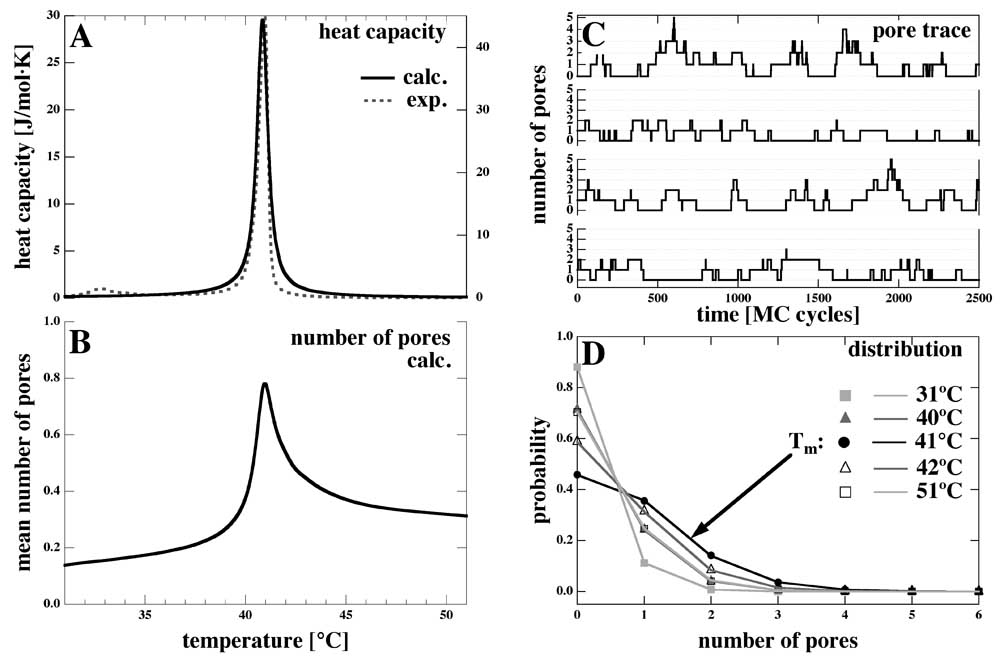}
	\parbox[c]{8cm}{ \caption{\small\textit{Monte Carlo simulations of a DPPC membrane. Top, left: Calculated heat capacity profile compared to the experimental profile of DPPC large unilamellar vesicles. Bottom, left: Calculated temperature dependence of the membrane permeability (mean number of pores). Top, right: Four traces of the number of pores (corresponding to conductance) calculated at the melting temperature, T$_m$, at 41$^{\circ}$C. Bottom, right: Probability distribution of the number of pores. The symbols originate from the simulation, the solid lines are fits by a Poisson-distribution. Simulation parameters are given in the text.
}
	\label{Figure4}}}
    \end{center}
\end{figure}

Fig. \ref{Figure4} shows the results of a simulated DPPC membrane. The simulated heat capacity profiles are compared to an experimental profile of DPPC large unilamellar vesicles in Fig. \ref{Figure4}A. The close agreement  of the two curves is not accidental but a consequence of using the parameters $\Delta H$, $\Delta S$ and the half width of the experimental heat capacity profile. Fig. \ref{Figure4}B shows the calculated mean number of pores in the simulation box, which is proportional to the conductance of the membrane. They show that the mean number of simulated pores reaches a maximum close to the transition maximum at 41$^{\circ}$C. This behavior is exactly that found by Blicher et al. \cite{Blicher2009} for dye permeation through large unilamellar vesicle membranes and confirms the concept of Nagle \& Scott \cite{Nagle1978b} that makes use of the maximum of the lateral compressibility at $T_m$. Fig. \ref{Figure4}C shows representative traces of the pore number computed at the melting temperature, $T_m$. These traces strongly resemble those from the BLM experiments (Figs. \ref{Figure2}),  \ref{Figure3a}) and  \ref{Figure3b}). Fig. \ref{Figure4}D shows the probability distribution of the number of pores for five different temperature below, at and above the melting temperature. The solid lines in this panel represent Poisson-distributions indicating that the calculated behavior matches the experimental one  (Fig. \ref{Figure3a}, bottom right). The Poissonian statistics is expected when pores form independent of each other.
\begin{figure}[htb!]
    \begin{center}
	\includegraphics[width=8.5cm]{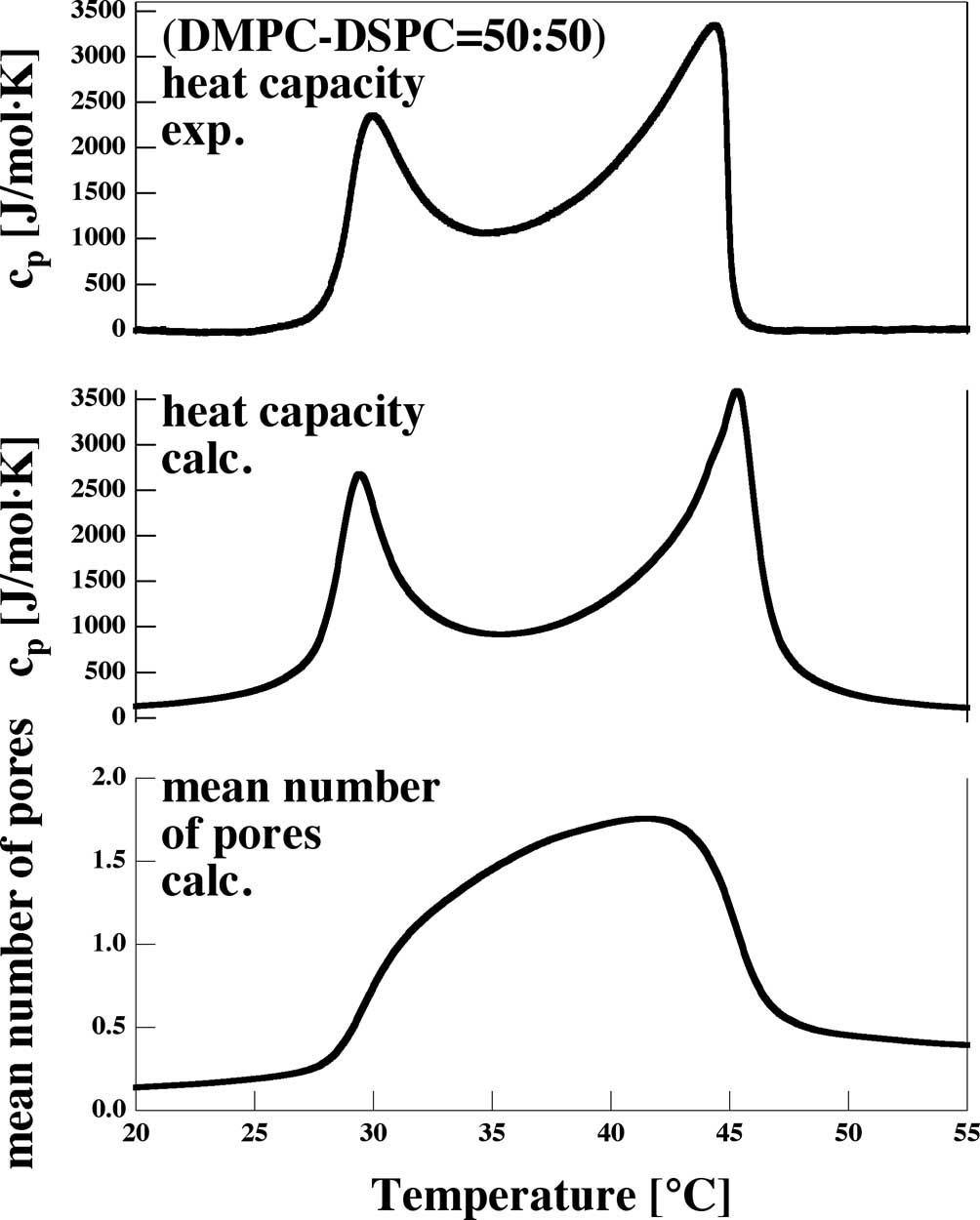}
	\parbox[c]{8cm}{ \caption{\small\textit{DMPC-DSPC mixture. Top: Experimental heat capacity profile. Middle: Calculated heat capacity profile from MC simulation. Bottom: Calculated mean number of pores as a function of temperature. One can recognize the the membrane permeability reaches a maximum in the chain melting regime. Simulation parameters are given in the text.}
	\label{Figure5}}}
    \end{center}
\end{figure}
The lipid-pore interaction parameters ($\omega_{pf}$ and $\omega_{pg}$) are given in the Methods section and have been chosen such that the absolute number of pores is of similar order as in the BLM experiment. Since the simulated matrix has about 10000 lipids while the black lipid membrane displays a much larger area of about $1.6\cdot 10^{10}$ lipids, the pore-lipid interaction parameters would have to be about 2-3 times higher if the simulation matrix had the same size than the BLM membrane. With such parameters, however, we would hardly ever see a conduction event in the simulation at a matrix size of 10000 lipids. Therefore, we have chosen smaller parameters. Qualitatively, however, the behavior of the pore formation rate with a maximum at $T_m$ is maintained when using the smaller model parameters given above. \\
The above concept also works in mixtures of different lipids with a broad melting peak. Fig. \ref{Figure5} shows the mean number of pores, i.e., the permeation profile, for a DMPC-DSPC= 50:50 lipid mixture. The corresponding experimental and simulated heat capacity profile is shown in the top panels of Fig. \ref{Figure5}. The maximum permeability ($\equiv$ mean number of pores) is found in the chain melting regime between the two outer $c_p$ maxima. \\
\begin{figure}[htb!]
    \begin{center}
	\includegraphics[width=8.5cm]{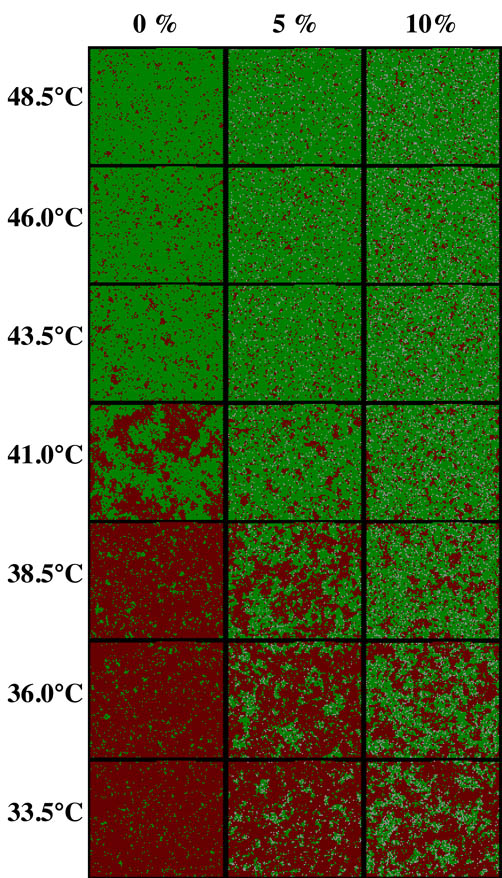}
	\parbox[c]{8cm}{ \caption{\small\textit{Simulations of DPPC membranes in the presence of anesthetics as a function of temperature and anesthetics concentration. Left column: Monte Carlo snapshots in the absence of anesthetics. Center column: In the presence of 5 mol\% anesthetics. Right column: In the presence of 10 mol\% anesthetics. The rows indicate different temperatures between 33.5$^{\circ}$C  and 48.5 33.5$^{\circ}$C. Gel lipids are given in red, fluid lipids in green, anesthetics molecules in white and pores in black. One can recognize that the presence of anesthetics shifts the melting temperature towards lower temperatures. Simulation parameters are given in the text. }
	\label{Figure6}}}
    \end{center}
\end{figure}

\textbf{The effect of anesthetics}
The famous Meyer-Overton rule \cite{Overton1901} states that the effectiveness of anesthetics is exactly proportional to their solubility in lipid membranes and independent of their chemical nature. Further, anesthetics are known to lower the phase transition temperature of lipid membranes. In a recent paper we have shown that the effect of anesthetics on pure lipid membranes can be well described by the well-known freezing-point depression law $T_m=\frac{RT_m^2}{\Delta H}\cdot x_A$, where $x_A$ is the molar fraction of anesthetics. The derivation of this formula is based on only two assumption: the anesthetics molecule is ideally soluble in the liquid phase, and that it is insoluble in the solid phase. In our simulation the anesthetics molecules are modeled as resembling lipids that occupy lattice sites. The perfect solubility of anesthetics in the fluid membrane phase can easily be modeled by assuming that the anesthetics-fluid interaction is zero ($\omega_{af}=0$ ), while the insolubility in the gel phase means that the anesthetics-gel parameter ($\omega_{ag}$ ) is very large. \\
\begin{figure}[hb!]
    \begin{center}
	\includegraphics[width=8.3cm]{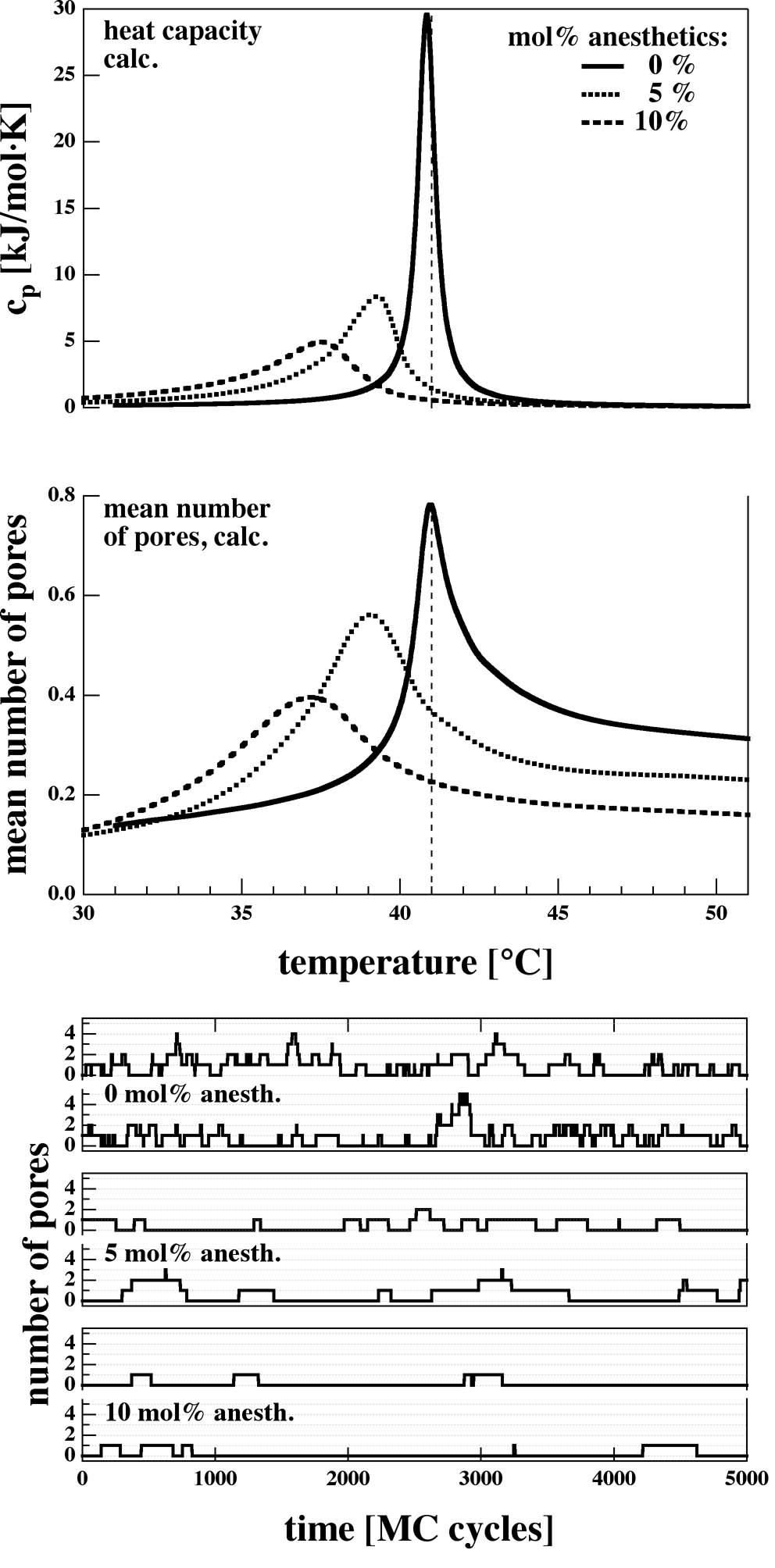}
	\parbox[c]{8cm}{ \caption{\small\textit{Monte Carlo Simulation of DPPC membranes in the presence of 0 mol\%, 5 mol\% and 10 mol\% of anesthetics in the membrane. Top:  Calculated heat capacity profiles show the shift of the melting events towards lower temperatures in agreement with the snapshots in Fig. 7. Middle: The calculated number of pores (corresponding to a permeability) follows the heat capacity profiles. The maximum permeability shifts towards lower temperatures upon addition of anesthetics. Bottom: Calculated traces of the number of pores indicate that the number of pore opening events decreases in the presence of anesthetics. Two traces for each anesthetics concentration are given. Simulation parameters are given in the text.}
	\label{Figure7}}}
    \end{center}
\end{figure}

Fig. \ref{Figure6} shows Monte-Carlo snapshot for DPPC membranes at seven different temperatures evenly spaced with 2.5 degrees. The three columns are calculated for 0mol\%, 5 mol\% and 10 mol\% anesthetics, respectively. Here, each anesthetics molecule occupies one lattice site. The melting temperature is again $T_m=41^{\circ}$C. In this figure red and green dots represent gel and fluid lipids, while white dots represent anesthetics molecules and black dots are pores. One can see how the increasing amount of anesthetics lowers the melting events as evident from the lower temperatures where one finds red gel domains. These simulations are further analyzed in Fig. \ref{Figure7}. Fig. \ref{Figure7} (top) shows the heat capacity profiles for the simulations in Fig. \ref{Figure6}. They are found to be very similar to those found experimentally in the case of DPPC with octanol \cite{Heimburg2007c} and show the lowering of the melting temperature in the presence of anesthetics. Note, that the critical anesthetic dose for tadpoles corresponds to 2.6 mol\%  of anesthetics in the membrane \cite{Heimburg2007c}, which is somewhat lower than the concentration used in this work. Fig. \ref{Figure7} (middle) shows the corresponding permeation profiles that show a similar shift towards lower temperatures in the presence of anesthetics. Fig. \ref{Figure7} (bottom) shows pore formation traces for three different anesthetics concentrations calculated at 41 $^{\circ}$C. These traces correspond to the experimental situation in Fig. \ref{Figure3a}. The mean number of pores is reduced. Since Monte-Carlo simulations are based on equilibrium thermodynamics the Monte-Carlo time scales of these traces do not correspond to a real time scale. This means that one cannot conclude from the simulation that the mean life time of the pore openings in the presence of anesthetics is longer as in the absence of anesthetics even though the traces seem to suggest that.

\section{Discussion}\label{Discussion}
In this paper we investigate the lipid ion channel formation in pure lipid membranes in the absence and the presence of anesthetics. The goal was to demonstrate the origin of the pore formation from the cooperative area fluctuations. We studied pore formation both experimentally by the Montal-M\"uller \cite{Montal1972} black lipid membrane technique, and by Monte-Carlo simulations. We show that pure lipid membranes in the absence of proteins display conduction events that resemble those seen in the presence of protein ion channels. For this reason we call them lipid ion channels. Such events have been shown before by a number of authors \cite{Yafuso1974, Antonov1980, Boheim1980, Kaufmann1983a, Kaufmann1983b, Gogelein1984, Antonov1985, Yoshikawa1988, Kaufmann1989c,  Woodbury1989, Antonov2005, Blicher2009}. Although there are obviously many reports about such events it is surprising that such behavior is not better known. While the pores in lipid membranes are shown to exist in the absence of proteins, the conduction of ions through proteins cannot be measured in the absence of lipid membranes. Since the events are so similar to the protein ion channel recordings one requires an unambiguous criterion for how to distinguish the events originating from proteins and lipid membranes. To our knowledge, no such criterion exists. Possibly, site-specific mutations of proteins related to changes in channel activity can be considered as an indirect criterion, but this may not be satisfactory because this only demonstrates that the protein are somewhat involved in the conduction process (cf. last paragraph of discussion).\\
The importance of phase transitions in lipid membranes for the permeability has been pointed out before \cite{Papahadjopoulos1973, Nagle1978b, Antonov1980, Cruzeiro1988, Corvera1992, Blicher2009}. The cooperative fluctuations in area render the pore formation more likely because the lateral compressibility is higher. This effect resembles the phenomenon of critical opalescence described by Einstein \cite{Einstein1910}. He calculated the work to move one molecule from one part of the volume to another in in order to obtain the density fluctuations. In a similar manner, the increased fluctuations in the 2D membrane render it easier to move lipids such that pores can be created. In the present paper we used this concept to describe the pore formation process by a simple statistical thermodynamics model in which lipids are moved while simultaneously reducing the area of the surrounding lipids. Since the change of the intensive thermodynamic variables (e.g., pressure, lateral pressure, chemical potentials of protons, of calcium or of anesthetic drugs) influences the melting behavior of membrane and thus the fluctuations in the membrane, it is straight forward to explore how changes in these variables influence permeability. For instance, it has been shown that voltage influences the phase transition temperature \cite{Antonov1990, Lee1995}. As evident in our measurements, the mean conductance through pores in lipid mixtures is linear in the range between about -150 to +150mV. For $|V_m|>150mV$ the current-voltage relation becomes non-linear indicating that high voltage influences the state of the membrane and induces more conduction events. It has further been shown that changes in calcium concentration  \cite{Gogelein1984, Antonov1985} and pH changes \cite{Kaufmann1983a, Kaufmann1983b} influence to occurrence of lipid ion channels. Here, we have used the chemical potential of anesthetics in the membrane as a variable. Our present work is continuation of a recent publication from our group \cite{Blicher2009} that shows how anesthetics influence the permeation process by influencing the melting process. Here, we have demonstrated this for the case of octanol and ethanol, which both shift transitions to a similar degree when the membrane concentration is the same. Both drugs lead to a reduction of the overall conductance.

Our simulations described the main feature found in experiments: The relation of the permeability and the heat capacity leading to maximum permeability at the heat capacity maximum. This confirms the main assumption of our study, which is that the cooperative area fluctuations are responsible for the occurrence of the lipid ion channels. It further describes quantitatively the effect of anesthetics on the heat capacity of the membrane. The anesthetics following the Meyer-Overton correlation also generate freezing-point depression. The simulation accurately describes this and therefore also describes the influence of anesthetics on the permeation. However, our pore formation model made the simplifying assumption that a pore has the size of one lipid. This is a reasonable approximation since we found in a recent paper that the conductance through the membrane is well described by assuming an aqueous pore of a diameter of 0.7nm, which is exactly the size of one lipid. The measurements in the presence of anesthetics show, however, that the pore size itself can vary and that it is different in the presence and absence of octanol and ethanol. This effect is not contained in our simulations.\\
The present Monte Carlo simulation is a coarse grain statistical thermodynamics model. Such models have successfully been used to describe cooperative phenomena in membranes \cite{Doniach1978, Mouritsen1983, Sugar1994, Heimburg1996a, Sugar1999, Ivanova2001}. Typically, the number of states for each lipid is reduced to two (Ising model) or 10 (Pink model). Naturally, this is only a very approximate description on the molecular scale. The reason why the lattice models work so nicely is that they are designed to describe cooperative behavior of many molecules - in our case the formation of domains and the cooperative fluctuations. Microscopic details of single lipids and pores described here by lattice sites are not expected to be accurate - and also to be of small relevance. As mentioned above, our simulation is designed such that a pore has always the size of one lattice site. The experiment on octanol shows that the pores in the presence of anesthetics are smaller than in the absence (currents at identical voltage are $\approx$ 90pA in the absence of octanol and only about $\approx$15pA in the presence of 15.9 mol\% octanol). Qualitatively, however, the simulation describes the likelihood of pore formation and thereby permeation quite well. In a recent paper we have shown that the permeation of fluorescence dyes through lipid vesicles is within experimental accuracy proportional to the heat capacity \cite{Blicher2009}. This paper also shows the quantized conduction events in BLMs that could be abolished by addition of octanol. Our simulations reproduce this close relation between heat capacity and conductance - and is consistent with the quantized nature of the conduction process.

It is likely that many other variable changes would influence the occurrence of lipid ion channels, e.g., proteins that associate or insert into membranes. It is known that band 3 protein of erythrocytes increases the melting temperature of model membranes \cite{Morrow1986}. Cytochrome b$_5$, on the other hand lowers transitions \cite{Freire1983}. Cytochrome c binds peripherally and increases the transition temperature \cite{Heimburg1996a}. Depending onwhether the experimental temperature is found below or above the melting temperature, one would expect that the respective proteins can either increase or lower the likelihood of lipid ion channel formation even if the proteins are not  known to form ion channels.\\
The mechanism of anesthesia has been under much debate and is unexplained until today. Until the mid 1970's lipid models were most popular because they can be easily related to the famous Meyer-Overton rule \cite{Overton1901} that states that the effectiveness of anesthetics and their partition coefficient are linearly related, independent of the chemical nature of the anesthetics molecules. Ever since the development of the patch clamp methods and the study of protein ion channel events, a lot of attention has been on the binding of anesthetics to receptors. This is due to the finding that ion channel events that have been attributed to proteins are influenced by anesthetics. Unfortunately, such interaction with proteins is not generally in agreement with the Meyer-Overton correlation. Further, we have shown here and in \cite{Blicher2009} that the finding of quantized events does not prove the action of a protein. Further, the influence of anesthetics on the conduction events does not prove a specific binding of the anesthetic drug to a receptor. We have shown here that octanol and ethanol influence the channel events even in the complete absence of proteins.\\
 In this context it is tempting to compare the effect of alkanols on protein ion channels with our findings. The influence of octanol on current fluctuations of lipid ion channels can be compared with those of neuronal nicotinic acetylcholine receptor (AChR, see, e.g., \cite{Zuo2004b}), which has been used for many years as a model to study the influence of alcohols \cite{Bradley1980, Bradley1984, Forman1995, Forman1997, Forman1999, Wood1991, Wu1994}. It has been shown that neuronal activity of nicotinic acetylcholine receptor (AChR) induced channels under the influence of octanol is markedly inhibited \cite{Zuo2004b}, which was observed as a decrease in current fluctuation amplitude from those channels. Ethanol, on the other hand was found not to be neutral for many protein ligand-gated ion channels such as N-methyl-D-aspartate (NMDA), serotonin (5-HT(3)), glycine and GABA receptors \cite{Harris1995, Harris1999}. The inhibitory action of ethanol was also observed in voltage-gated Ca2+ channels \cite{Mullikin1994} and very recently also in potassium channels, which were seen as very important targets of ethanol   \cite{Brodie2007}. Similar findings were reported for experiments with AChR channels subjected to 100mM of ethanol \cite{Zuo2004}. The above reports demonstrate that the effect of octanol and ethanol on membrane protein receptors is quite similar to that found by us for the effect of these alcohols on lipid ion channels. In this context it is also interesting to note that protein ion channels have been shown to respond to the phase behavior of lipids \cite{Cannon2003} or to different lipid environments \cite{Schmidt2006}.\\


\small{

}
\end{document}